\documentstyle[pre,aps]{revtex}
\begin{document}
\title{Solidity of Viscous Liquids}
\twocolumn
\author{Jeppe C. Dyre}
\address{Department of Mathematics and Physics (IMFUFA), 
Roskilde University, P.O.Box 260, DK-4000 Roskilde, Denmark}
\date{\today}
\maketitle{}

\begin{abstract}

Recent NMR experiments on supercooled toluene and glycerol by
Hinze and B{\"o}hmer show that small rotation angles dominate
with only few large molecular rotations.
These results are here interpreted by assuming that viscous
liquids are solid-like on short length scales.
A characteristic length, the ``solidity length'', separates
solid-like behavior from liquid-like behavior.

\end{abstract}


The viscosity of a liquid approaching the glass transition
\cite{kau48,har76,bra85,joh85,jac86,ang88,sch90,ang91,hun93,%
2ang95,moh95,edi96} is typically a factor $10^{15}$ larger than
the viscosity of ordinary liquids like room-temperature water or
ethanol. 
Although viscosity is just a parameter entering the
Navier-Stokes equation believed to describe all liquids close to
equilibrium, this enormous difference raises the question:
Are viscous liquids {\it qualitatively} different from ordinary
liquids or is the difference just {\it quantitative}?
Below, it is argued that the former is the case. 
The idea is that viscous liquids behave like solids on short
length scales.
It is shown that this leads to a prediction consistent with the
results of recent NMR experiments by Hinze and B{\"o}hmer
\cite{hin98,boh98}. 

In many phenomenological models of viscous liquids
\cite{bra85,gib58,%
coh59,ada65,gol69,wil75,don81,sti88,cha93,dyr95,kiv96}
flow proceeds via sudden reorientations of molecules, ``flow
events'', which are rare because of the large energy barriers to
be overcome \cite{kau48,ada65,gol69,sti88}.
Kauzmann referred to flow events as ``jumps of molecular units
of flow between different positions of equilibrium in the
liquid's quasicrystalline lattice'' \cite{kau48}.
It is this point of view that is explored here:
Most molecular motion is purely vibrational and in the time
between two flow events a viscous liquid is in a state of
elastic equilibrium, just like a solid.
However, elastic equilibrium only persists on a certain length
scale beyond which the liquid does not display solid-like
behavior (this point is returned to below). 

Recently, Hinze and B{\"o}hmer studied reorientation of toluene
and glycerol molecules by means of two-dimensional time-domain
NMR spectroscopy \cite{hin98,boh98}.
The rotation angle distribution is dominated by small angles with
a small, but significant fraction of larger rotation angles.
These findings were interpreted as follows \cite{boh98}.
The large-angle rotations are those required to cross a local
energy barrier.
Upon barrier crossing local strains are created.
These strains are relaxed through small positional and angular
adjustments, not only by the molecules in the immediate vicinity
but also by those further away.
Briefly, large-angle rotations are ``causes'' and small-angle
rotations are ``effects''.
Accepting this picture, we now proceed to show that the rotation
angle distribution for small angles may be derived from the
fact that viscous liquids have slow density fluctuations,
assuming these liquids are solid-like on short length scales.

Consider first density fluctuations.
Viscous liquids have long average relaxation times (roughly
proportional to viscosity according to the Maxwell relation).
These long relaxation times are basically the time between two
flow events involving the same molecule.
Not only enthalpy or shear stress relaxes on this time scale,
but so does density.
This has been known for many years from the fact that glass has
smaller compressibility than corresponding equilibrium viscous
liquid.
More recently, measurements of the frequency-dependent bulk
modulus of viscous liquids \cite{chr94} revealed a loss peak
around the inverse of the Maxwell relaxation time; via the
fluctuation-dissipation theorem this shows directly that there
are slow density fluctuations.
Slow density fluctuations in viscous liquids lead to ``dynamic
heterogeneities'', a major research topic in the 1990's
\cite{per96}.
Dynamic heterogeneities have been observed, e.g., in light
scattering experiments \cite{moy93,wan96}, NMR experiments
\cite{boh96}, time resolved optical spectroscopy \cite{cic95},
and computer simulations \cite{hur95}.

As mentioned, slow density fluctuations take place on a
time-scale basically determined by the rate of flow events.
A flow event is a rapid reorientation of molecules, probably
lasting just a few picoseconds.
After a flow event the molecules involved have different
relative orientations.
Generally, density changes somewhat at the place of a flow
event (this is the cause of slow density fluctuations).
As a simple model, assume isotropic flow events involving
molecules confined to a sphere of radius $r_0$ before the flow
event.
We now proceed to calculate the rotation angle probability
distribution for small angles.
The induced movement of the surroundings is calculated by means
of solid elasticity theory. 
If the change of radius is $\Delta r$, the displacement of the
surroundings is given \cite{l+l,wyl80,dyr98} by $u_r=\Delta r
(r/r_0)^{-2}$, where $r$ is the distance to the flow event.
The average rotation angle $\phi$ is proportional to the strain
tensor, which in turn is formed from first order derivatives of
$u_r$. 
Consequently, $\phi\propto r^{-3}$ (a detailed calculation
gives $\langle \phi^2\rangle=(6/5)r_0^4 (\Delta r)^2 r^{-6}$).
The rotation angle probability distribution is given by
$P(\phi)=P(r)|dr/d\phi |$, where $P(r)\propto r^2$ from geometry.
Since $|dr/d\phi |\propto r^4$ we thus find $P(\phi)\propto r^6$.
Thus, [for small rotation angles] $P$ is given by

\begin{equation}\label{1}
P(\phi)\ \propto\ \phi^{-2}\,.
\end{equation}
Presently, it is not possible to determine $P(\phi)$ accurately
from experiments.
It should be noted, though, that since $\sin(\phi)\simeq\phi$ for
small $\phi$, Eq. (\ref{1}) is consistent with the rotation angle
distribution tentatively inferred from NMR experiments on
glycerol, $P(\phi)\propto 1/\sin^2(\phi)$ \cite{boh98}.

The rotation angle distribution Eq. (\ref{1}) is not
normalizable, reflecting the fact that in the above derivation
all molecules of the liquid rotate slightly following a single
flow event.
This, however, is not realistic; there is a ``solidity length'' 
$l$, beyond which flow events effectively do not induce molecular
rotations.
To estimate $l$ note that elastic displacements propagate with
the velocity of sound, $c$.
Consider a sphere with radius $R$.
Within this sphere there are $N=(R/r_0)^3$ possible locations for
flow events.
A molecule at the center of the sphere only ``feels'' the full
effects from any flow event in the sphere if the following
condition is obeyed:
The displacement deriving from such a flow event must propagate 
throughout the sphere and elastic equilibrium be reestablished
before the next flow event occurs.
If $\tau$ is the average relaxation time, the average time
between two flow events within the sphere is 
$\tau /N=\tau(R/r_0)^{-3}$.
This time must be longer than or equal to $R/c$.
To estimate the solidity length $l$ we use equality for $R=l$ and
note that $c$ is, of course, the sound velocity of the glassy
state, $c_{\rm glass}$.
This leads to

\begin{equation}\label{2}
l^4\ =\ r_0^3\ \tau\ c_{\rm glass}\,.
\end{equation}
The solidity length diverges slowly as $\tau\rightarrow\infty$.
To get a feeling of the order of magnitude of $l$, consider the
case where $\tau = 1 {\rm s}$.
Assuming $r_0= 5 {\rm{\AA}}$ and $c_{\rm glass}=10^3 {\rm m/s}$
one finds $l\simeq 6,000 {\rm{\AA}}$.

To conclude, below the glass transition, of course, viscous
liquids are solid for all practical purposes.
However, it has been argued here that even above the glass
transition viscous liquids in certain respects behave more like
solids than like less-viscous liquids.
The solid-like behavior takes place on length scales below the
solidity length $l$.
Note that $l$ diverges when $\tau$ diverges; however $l$ is
probably unrelated to the Adam-Gibbs characteristic length that
also diverges with $\tau$ \cite{ada65}.
Rather, the solidity length $l$ is similar to the length scale
``related to solid-like behavior'' recently discussed by
Ahluwalia and Das within ideal mode-coupling theory \cite{ahl98},
a length scale representative of the distance over which the
liquid has enough structure to sustain propagating shear waves.
However, the exact relation between the length discussed by
Ahluwalia and Das and the solidity length introduced here remains
to be determined.

\acknowledgements
 This work was supported by the Danish Natural Science Research
Council.

\end{document}